 \documentclass[prb,reprint,amsmath,amssymb,showpacs]{revtex4-1}%
\usepackage{hyperref}
\usepackage{graphicx}
 \usepackage{amsmath}
\usepackage{graphicx}%
\usepackage{amsfonts}%
 \usepackage{amssymb}
\usepackage{color}
\usepackage{setspace}
\usepackage{textcomp}
\usepackage{makecell}
\usepackage{bm}

\begin{document}

\title{The induced permittivity increment of electrorheological fluids in an applied electric field in association with chain formation: A Brownian Dynamics simulation study}
\author{D\'avid Fertig}
\author{Dezs\H{o} Boda}
\email[Author for correspondence:]{boda@almos.vein.hu}
\affiliation{Center for Natural Sciences, University of Pannonia, Egyetem u. 10, Veszpr\'{e}m, 8200, Hungary}

\author{Istv\'{a}n Szalai}
\affiliation{Research Centre for Engineering Sciences, University of Pannonia, Egyetem u. 10, Veszpr\'{e}m, 8200, Hungary}
\affiliation{Institute of Mechatronics Engineering and Research, University of Pannonia, Gasparich M\'{a}rk u. 18/A, Zalaegerszeg, 8900, Hungary}

\date{\today}


\begin{abstract}
We report Brownian Dynamics simulation results for the relative permittivity of electrorheological (ER) fluids in an applied electric field. 
The relative permittivity of an ER fluid can be calculated from the Clausius-Mosotti (CM) equation in the small applied field limit.
When a strong field is applied, however, the ER spheres are organized into chains and assemblies of chains in which case the ER spheres are polarized not only by the external field but by each other.
This manifests itself in an enhanced dielectric response, e.g., in an increase in the relative permittivity. 
The correction to the relative permittivity and the time dependence of this correction is simulated on the basis of a model in which the ER particles are represented as polarizable spheres.
In this model, the spheres are also polarized by each other in addition to the applied field.
Our results are qualitatively similar to those obtained by Horváth and Szalai experimentally (\textit{Phys. Rev. E}, {86}, 061403, 2012). 
We report characteristic time constants obtained from bi-exponential fits that can be associated with formation of pairs and short chains as well as with aggregation of chains.
The electric field dependence of the induced dielectric increment reveals the same qualitative behavior that experiments did: three regions with different slopes corresponding to different aggregation processes are identified.  
\end{abstract}


\maketitle

\section{Introduction}

In electrorheological (ER) fluids~\cite{winslow_jap_1949} fine non-conducting solid particles are suspended in an electrically insulating liquid with the particles having larger relative permittivity than the solvent.
Then, an applied electric field induces polarization charges at the arising dielectric boundaries {that can be expressed as a multipole expansion with dipoles being the dominant terms}.

The interactions of these dipoles lead to a structural change in the ER fluid known as the ER response.
This structural change is the aggregation of ER particles first into shorter, then into longer chains due to the fact that the head-to-tail position of two dipoles along the direction of the applied field is a minimum-energy configuration.
In the case of strong applied fields, the chains form larger clusters, for example, columnar structures.

This structural change results in changes in major physical properties of the ER fluid.
The externally controllable, fast and reversible change in viscosity, for example, makes ER fluids a central component of various devices, such as brakes, clutches, dampers, and valves \cite{Duclos_1992,havelka_progress_ER_1995}. 

Another physical quantity whose change can be relatively easily tracked by measuring the change in the capacitance of a measuring cell when the electric field is switched on is the relative permittivity, $\epsilon$.
Several experimental studies have been reported for the nonlinear dielectric properties of ER fluids.~\cite{adolf_l_1995,tao1994electrorheological,wen_pre_1997,wen_apl_1998,rzoska2006nonlinear} 
Horváth and Szalai~\cite{horvath_pre_2012} have proposed a new method that made the measurement of continuous changes in the increments in permittivity possible. 
They determined the field dependence of the change in dielectric permittivity
\begin{equation}
 \Delta \epsilon (E) = \epsilon(E) - \epsilon(E=0),
\end{equation}
with $t\rightarrow \infty$ and also the time dependence
\begin{equation}
 \Delta \epsilon (t) = \epsilon(t) - \epsilon(t=0),
\end{equation} 
where the electric field is switched on from 0 to $E$ at $t=0$.
{The time dependence and electric field dependence are shown in Fig.~3 and Fig.~4 of Ref.~\onlinecite{horvath_pre_2012}, respectively.}

They found that the time dependence can be described with a bi-exponential fit
\begin{equation}
 \Delta \epsilon(t) = A\left( 1-e^{-t/\tau_{1}} \right) + B\left( 1-e^{-t/\tau_{2}} \right).
 \label{eq:biexp}
\end{equation} 
They hypothesized that the time constant $\tau_{1}$ can be associated with the process of formation of pairs (and, perhaps, short chains), because $\tau_{1}$ is very similar to the characteristic time of pair formation derived from a kinetic rate theory.~\cite{baxter_drayton_jr_1996}
The time constant $\tau_{2}$, on the other hand, was heuristically corresponded to formation of long chains and their aggregation and proved to be an order of magnitude larger than $\tau_{1}$, $\tau_{2}\approx 10\tau_{1}$.

In this paper, we investigate how change in the dielectric permittivity is associated with structural changes (formation of chains of various lengths, and the aggregation of chains into columnar structures) by computer simulations.
Although several simulation studies have been reported for cluster formation~\cite{Klingenberg_1989,see_psj_1991,Toor_1993,hass_pre_1993,climent_langmuir_2004,dominguezgarcia_pre_2007}, order parameters~\cite{hass_pre_1993,tao_prl_1994,tao_ijmpb_1994,baxter_drayton_jr_1996,enomoto_physa_2002}, diffusion constant~ \cite{Klingenberg_1989,whittle_jnnfm_1990,hass_pre_1993}, pair distribution functions~\cite{whittle_jnnfm_1990,hass_pre_1993}, relaxation times~\cite{Heyes_1990,Toor_1993,hass_pre_1993,cao_jpcb_2006}, aggregation kinetics~\cite{see_psj_1991}, and stress under shear~\cite{Heyes_1990,whittle_jnnfm_1990,bonnecaze_jcp_1992,baxter_drayton_jr_1996,cao_jpcb_2006}, we are not aware of any paper addressing the dielectric properties of ER fluids.

The field dependence revealed three regimes with different slopes of the $\Delta \epsilon$ vs.\ $E$ function (Fig.~4 of Ref.~\onlinecite{horvath_pre_2012}).
Horv\'ath and Szalai hypothesized that at low electric fields (below a threshold value $E_{1}$) the induced dipoles are not strong enough to generate chain formation and $\Delta \epsilon$ increases with $E$, because the probability of the ER spheres to approach each other becomes larger.
Above $E_{1}$ chain formation begins with a large slope, because the parts of a chain can find one another easier at large electric fields.
Above the threshold value $E_{\mathrm{h}}$ the chains start to form columnar structures.
This process is less accelerated by the large electric field, because chains can also repulse each other when they are not in the appropriate mutual configuration with respect to each other.
Here, we support this hypothesis with computer simulation results.

\section{Models and methods}
\label{sec:model}

The ER fluid is modeled as monodisperse dielectric spheres of relative permittivity $\epsilon_{\mathrm{in}}$ inside the sphere immersed in a carrier liquid of relative permittivity $\epsilon_{\mathrm{out}}$.
The radius of the spheres is $R$, while their diameter is $d=2R$.
If an electric field, $\mathbf{E}$ is applied on the sphere, a polarization charge density is induced on the surface of the sphere that can be approximated with an ideal point dipole placed in the center of the sphere computed as \cite{jackson}
 \begin{equation}
  \bm{\mu}=4\pi \epsilon_{0} \left( \frac{\epsilon_{\mathrm{in}}-\epsilon_{\mathrm{out}}}{\epsilon_{\mathrm{in}}+2\epsilon_{\mathrm{out}}} \right) R^{3}\mathbf{E} = \alpha \mathbf{E}, 
  \label{eq:dipole}
 \end{equation} 
 where 
 \begin{equation}
 \alpha = 4\pi \epsilon_{0} \left( \frac{\epsilon_{\mathrm{in}}-\epsilon_{\mathrm{out}}}{\epsilon_{\mathrm{in}}+2\epsilon_{\mathrm{out}}} \right) R^{3} 
 \label{eq:alpha}
 \end{equation} 
is the particle polarizability, $E=|\mathbf{E}|$, and $\epsilon_{0}$ is the permittivity of vacuum.
If it is further assumed that the characteristic time of the rearrangement of the surface charge during the movement of the particles is much smaller than the characteristic time of the rotation of the ER particles, the $\bm{\mu}$ dipole always points into the direction of $\mathbf{E}$ even if the sphere rotates.

If we take a system of $N$ particles at positions $\{\mathbf{r}_{j}\}$, the electric field exerted on dipole $i$ by dipole $j$ is
\begin{equation}
\mathbf{E}_{j}(\mathbf{r}_{i}) = \frac{1}{4\pi \epsilon_0} \frac{3\mathbf{n}_{ij}(\mathbf{n}_{ij}\cdot \bm{\mu}_{j})-\bm{\mu}_{j}}{r_{ij}^{3}},
\label{eq:field}
\end{equation} 
where $r_{ij}=|\mathbf{r}_{ij}|$ and $\mathbf{n}_{ij}=\mathbf{r}_{ij}/r_{ij}$ with $\mathbf{r}_{ij}=\mathbf{r}_{i}-\mathbf{r}_{j}$.

In Eq.~\ref{eq:dipole}, the electric field at $\mathbf{r}_{i}$ is a sum of the applied field, $\mathbf{E}^{\mathrm{appl}}$ (points to the $z$ direction), and the electric field produced by all the other dipoles, $\mathbf{E}(\mathbf{r}_{i})=\sum_{j\ne j} \mathbf{E}_{j}(\mathbf{r}_{i})$.
The total dipole moment
\begin{equation}
 \bm{\mu}^{\mathrm{tot}}_{i} = \alpha \mathbf{E}^{\mathrm{appl}} + \alpha \mathbf{E}(\mathbf{r}_{i}) = \bm{\mu}^{\mathrm{appl}}_{i} + \bm{\mu}^{\mathrm{part}}_{i}
 \label{eq:iterate_mu}
\end{equation} 
then is induced by these two components and is split into the terms $\bm{\mu}^{\mathrm{appl}}_{i}$ and $\bm{\mu}^{\mathrm{part}}_{i}$ accordingly.
The dipole moment $\bm{\mu}_{i}^{\mathrm{appl}}$ induced by the applied field is constant, while the dipole moment $\bm{\mu}_{i}^{\mathrm{part}}$ induced by all the other ER particles needs to be calculated by an iterative procedure.~\cite{vesely_jcompp_1977} 

{In the rheological literature, it is usual to ignore the polarization of the particles by each other ($\bm{\mu}^{\mathrm{part}}_{i}=0$). 
There are, however, important exceptions.~\cite{tao_prl_1991,tao_pre_1993,tao_prl_1994,bonnecaze_jcp_1992,wang_ijmpb_1996,wang_cpl_1997,martin_jcp_1998,hynninen_prl_2005}}
In this work, we present the full self consistent solution of Eqs.~\ref{eq:field}--\ref{eq:iterate_mu}.

If we introduce the force exerted on dipole $\bm{\mu}_{i}$ by dipole $\bm{\mu}_{j}$ (irrespective whether they are induced by $E^{\mathrm{appl}}$ or by other particles) is
\begin{eqnarray}
 \mathbf{f}^{\mathrm{dip}}_{ij}(\mathbf{r}_{ij},\bm{\mu}_{i},\bm{\mu}_{j}) = - (\bm{\mu}_{i}\cdot \nabla_{i}) \mathbf{E}_{j}(\mathbf{r}_{i}) = \nonumber \\
=  \frac{1}{4\pi \epsilon_0} 
\frac{1}{r_{ij}^{4}} \left\{ 3 \left[ 
\bm{\mu}_{i}  (\mathbf{n}_{ij}\cdot \bm{\mu}_{j}) +
\bm{\mu}_{j}  (\mathbf{n}_{ij}\cdot \bm{\mu}_{i}) + \right. \right. \nonumber \\  
+ \left. \left. \mathbf{n}_{ij} (\bm{\mu}_{i} \cdot \bm{\mu}_{j}) \right]
- 15 \mathbf{n}_{ij} (\mathbf{n}_{ij}\cdot \bm{\mu}_{i}) 
                     (\mathbf{n}_{ij}\cdot \bm{\mu}_{j}) 
\right\}  
\label{eq:fdd},
\end{eqnarray} 
we can express the force exerted on dipole $\bm{\mu}^{\mathrm{appl}} _{i}$ by dipole $\bm{\mu}^{\mathrm{appl}} _{j}$ as
\begin{equation}
 \mathbf{f}_{ij}^{\mathrm{appl}}  = \mathbf{f}^{\mathrm{dip}}_{ij}(\mathbf{r}_{ij},\bm{\mu}^{\mathrm{appl}} _{i},\bm{\mu}_{j}^{\mathrm{appl}} ) 
\end{equation} 
and the force exerted on dipole $\bm{\mu}^{\mathrm{appl}} _{i}$ by dipole $\bm{\mu}^{\mathrm{part}} _{j}$ as
\begin{equation}
 \mathbf{f}_{ij}^{\mathrm{part}} = \mathbf{f}^{\mathrm{dip}}_{ij}(\mathbf{r}_{ij},\bm{\mu}^{\mathrm{appl}} _{i},\bm{\mu}_{j}^{\mathrm{part}}) .
\end{equation} 
The finite size of the ER particles is taken into account with a short-range repulsive core potential for wich the cut \& shifted Lennard Jones (LJ) potential also known as the Weeks-Chandler-Anderson (WCA) potential is used. 
The WCA force is defined as
\begin{equation}
 \mathbf{f}_{ij}^{\mathrm{WCA}}(\mathbf{r}_{ij}) = \left\{
 \begin{array}{ll}
  \mathbf{f}_{ij}^{\mathrm{LJ}} (\mathbf{r}_{ij})  \; &\mathrm{if} \; \; r_{ij}<r_{\mathrm{c}} \\
  0 \; &\mathrm{if} \; \; r_{ij}>r_{\mathrm{c}}
 \end{array}
\right. ,
\label{eq:fcslj}
\end{equation} 
where 
\begin{equation}
 \mathbf{f}_{ij}^{\mathrm{LJ}} (\mathbf{r}_{ij}) = 24 \varepsilon^{\mathrm{LJ}} \left[2 \left( \frac{d}{r_{ij}} \right)^{12} - \left( \frac{d}{r_{ij}} \right)^{6} \right] \frac{\mathbf{r}_{ij}}{r_{ij}^{2}}
\end{equation} 
is the LJ force ($r_{\mathrm{c}}=2^{1/6}d$ is the cutoff distance at the minimum of the LJ potential).

The trajectories of the particles can be computed from Langevin's equations of motion \cite{lemons_1997}
\begin{equation}
m\frac{d\mathbf{v}_{i}(t)}{dt} = \mathbf{F}_{i}\left(\mathbf{r}_{i}(t)\right) -m \gamma \mathbf{v}_{i}(t) + \mathbf{R}_{i}(t),
\label{eq:langevin}
\end{equation} 
where 
\begin{equation}
\mathbf{F}_{i} = \sum_{j} (\mathbf{f}_{ij}^{\mathrm{WCA}} + \mathbf{f}_{ij}^{\mathrm{appl}}  + \mathbf{f}_{ij}^{\mathrm{part}})
\end{equation} 
is the systematic force, $-m \gamma \mathbf{v}_{i}(t)$ is the frictional force, $\mathbf{R}_{i}(t)$ is the random force, $\mathbf{r}_{i}$, $\mathbf{v}_{i}$, $m$, and $\gamma$ are the position, the velocity, the mass, and the friction coefficient of particle $i$, respectively. 
{The friction coefficient can be computed from Stokes' law as $ \gamma =  3\pi \eta d/m$.}

To solve this stochastic differential equation we use the GJF-2GJ version~\cite{jensen_mp_2019} of a collections of algorithms proposed by Gr{\o}nbech-Jensen and Farago:~\cite{Gronbech_Jensen_mp_2013,farago_physicaA_2019,jensen_mp_2019}
\begin{equation}
 v^{n+\frac{1}{2}} = a v^{n-\frac{1}{2}} + \frac{\sqrt{b}\Delta t}{m} f^{n} + \frac{\sqrt{b}}{2m} \left( R^{n}-R^{n+1} \right)
 \label{eq:farago_v}
\end{equation}
\begin{equation}
 r^{n+1}=r^{n} + \sqrt{b} v^{n+\frac{1}{2}}  \Delta t ,
 \label{eq:farago_r}
\end{equation}
where $r^{n}=r(t^{n})$ is any position coordinate of any particle, $v^{n}=v(t_{n})$ is any velocity coordinate of any particle, $t^{n}=n\Delta t$ is the time in the $n$th time step, $\Delta t$ is the time step, 
 $a = (1-\gamma \Delta t/2)/(1+\gamma \Delta t /2)$ ,
 $b =  (1)/(1+\gamma \Delta t /2)$,
$t_{n+\frac{1}{2}}=t_{n}+\Delta t/2$, and $t_{n-\frac{1}{2}}=t_{n}-\Delta t/2$.
The discrete time noise, $R^{n}$, is a random Gaussian number with properties $\langle R^{n} \rangle =0$ and $ \langle R^{m}R^{n}\rangle = 2 kT \gamma m \Delta t \delta_{mn}$ with $\delta_{mn}$ being the Kronecker-delta.

{The Brownian Dynamics simulations have been performed in a cubic simulation cell using periodic boundary conditions. 
The ensemble can be considered canonical because $V$ and $N$ are fixed, while the temperature is also constant because the system is thermostated by the Langevin integrator via the fluctuation--dissipation theorem.
The dipolar interactions were truncated at the half of the cell width. 
System size dependence has been analyzed in our previous work.~\cite{fertig_aipadv_2021}}


\section{Results and Discussion}
\label{sec:results}

\begin{table}[t]
\caption{Reduced quantities defined with $T$, $d$, $m$ or with $T$, $d$, $\rho_{\mathrm{in}}$.}
\label{tab:reduced}
 	\def\arraystretch{1.3}
 	\centering
\begin{tabular}{ll} \hline 
Quantity & Reduced quantity \\ \hline
Time & $t^{*}=t\sqrt{kT/md^{2}}=t\sqrt{6kT/\pi \rho_{\mathrm{in}}d^{5}}$ \\
Distance & $r^{*}=r/d$ \\
Density & $\rho^{*}=\rho d^{3}$ \\
Velocity & $v^{*}=v\sqrt{m/kT}=v\sqrt{\pi \rho_{\mathrm{in}}d^{3}/6kT}$ \\
Energy  & $u^{*}=u/kT$ \\
Force  &  $ F^{*}=Fd/kT$ \\
Electric field & $E^{*}=E \sqrt{4\pi \epsilon_0 d^{3}/kT}$ \\
Dipole moment  & $\mu^{*}=\mu/\sqrt{4\pi\epsilon_{0}kTd^{3}} $ \\
Polarizability  & $\alpha^{*} = \alpha /4\pi\epsilon_{0}d^{3}$ \\
Friction coefficient & $\gamma^{*} = \gamma \sqrt{md^{2}/kT} = \gamma \sqrt{\rho_{\mathrm{in}}d^{5}/6kT} $ \\ \hline
\end{tabular}
\end{table}

In this work, we use reduced units that are collected in Table \ref{tab:reduced}.
They express physical quantities as dimensionless numbers obtained by dividing a quantity in a physical unit by a unit quantity, $r^{*}=r/d$, for example. 
In addition to $T$ and $d$, we can use either $m$ or $\rho_{\mathrm{in}}$ (mass density of the ER particles) to define the reduced quantities, because the mass depends on $m$ and $d$ through $m=\rho_{\mathrm{in}}d^{3}\pi/6$.
The diffusion constant in the high coupling limit can be expressed by Einstein's relation:  $D=kT/m\gamma$.
The square of the reduced dipole moment
 \begin{equation}
  (\mu^{*})^{2} = \frac{\mu^{2}/4\pi \epsilon_{0}d^{3}}{kT} 
 \label{eq:squaredip}
  \end{equation} 
is an important parameter because it relates the ordering effect of the dipolar energy to the disordering effect of thermal motion.
It is proportional to the $\lambda$ parameter used in the literature.
If $(\mu^{*})^{2}$ is large, the dipolar interactions are strong enough to induce chain formation, while if it is small, thermal motion prevents chain formation.

In this work, our main goal is to study the dielectric response of the ER fluid and to analyze the relationship of this response to chain formation in the system.
The relative permittivity of an ER fluid can be computed from the corrected Clausius-Mosotti (CM) equation that can be derived from a polarization formula:~\cite{neumann_mp_1983}
\begin{equation}
 \dfrac{\epsilon -1}{\epsilon +2} =\dfrac{1}{3\epsilon_0} \dfrac{\left\langle P \right\rangle }{E^{\mathrm{appl}}} ,
\label{neumann}
\end{equation}  
where $P$ is the polarization density obtained from the sum of the dipoles $\mu^{\mathrm{appl}}=\alpha E^{\mathrm{appl}}$ induced directly by the external field and the average dipole moment $\left\langle \mu^{\mathrm{part}} \right\rangle$ induced by other particles:
\begin{equation}
\left\langle P \right\rangle = \frac{N\mu^{\mathrm{appl}} + N \left\langle \mu^{\mathrm{part}} \right\rangle}{V}  = \rho \mu^{\mathrm{appl}} \left( 1 + \frac{\left\langle \mu^{\mathrm{part}} \right\rangle}{\mu^{\mathrm{appl}}} \right) ,
\label{cCM2}
\end{equation} 
{here $V$ is the considered volume, and $N$ the number of particles in it.}
Eqs.\ \ref{eq:dipole}, \ref{neumann} and\ \ref{cCM2} result in the corrected CM equation
\begin{equation}
 \dfrac{\epsilon -1}{\epsilon +2} =\dfrac{\alpha \rho}{3\epsilon_0}  \left( 1 + \dfrac{\left\langle \mu^{\mathrm{part}} \right\rangle}{\mu^{\mathrm{appl}}} \right),
\end{equation}  
where $\rho=N/V$ is the number density, and the correction factor $\left\langle \mu^{\mathrm{part}} \right\rangle / \mu^{\mathrm{appl}}$ is the average induced dipole due to other particles normalized by the dipole due to the external field.
The quantity $\left\langle \mu^{\mathrm{part}} \right\rangle$ is directly provided by the simulations.
We have investigated the correction to the CM equation in the case of nonpolar fluids (e.g., carbon dioxide) by Monte Carlo simulations.~\cite{valisko_jcp_2009}

In the low-field-strength limit, the original CM equation is recovered (the correction factor is zero), because formally an ensemble of ER particles corresponds to an ensemble of non-polar, but polarizable molecules.
The CM equation is based on the Lorentz formula~\cite{lorentz} for the internal field and ignores the fact that a particle is also polarized by other particles not only by the external field.~\cite{keyes_pr_1931}

The number of particles is fixed at $N=256$ in our simulations.
This number proved to be sufficient on the basis of the system size analysis provided in our previous study.~\cite{fertig_aipadv_2021}

The value of the reduced friction coefficient is fixed at $\gamma^{*}=100$.
This value made simulations feasible, because the system developed relatively fast at the fixed value of the time step $\Delta t^{*}=0.005$.
The effect of larger values of $\gamma^{*}$ that are more characteristic of typical ER fluids is analyzed in our previous work.~\cite{fertig_aipadv_2021}
The reduced density is fixed at $\rho^{*}=0.05$.
The parameters that we change are the reduced electric field $E^{*}$ and the reduced polarizability $\alpha^{*}$.

To characterize time dependence, we show values of block averages (denoted by $\langle \dots \rangle_{\mathrm{b}}$ for various physical quantities as functions of $t^{*}$. 
In this work, we performed $M_{\mathrm{b}}{=}5000$ time steps in a block.
We perform $M_{0}=50,000$ time steps ($20$ blocks) in the absence of applied electric field ($E^{\mathrm{appl}}{=}0$), after which the electric field is instantaneously switched on. 
Then we performed  $M_{\mathrm{E}}=450,000$ time steps ($180$ blocks) in the presence of a constant applied field. 

Such a cycle was started over and done several times and averaged to smooth out noise.
When we start a cycle over, we restart from a freshly generated initial configuration in a completely disordered state without chains.
This way, the subsequent periods are independent and can be averaged.

\begin{figure}[t!]
	\centering
	\includegraphics[width=0.45\textwidth]{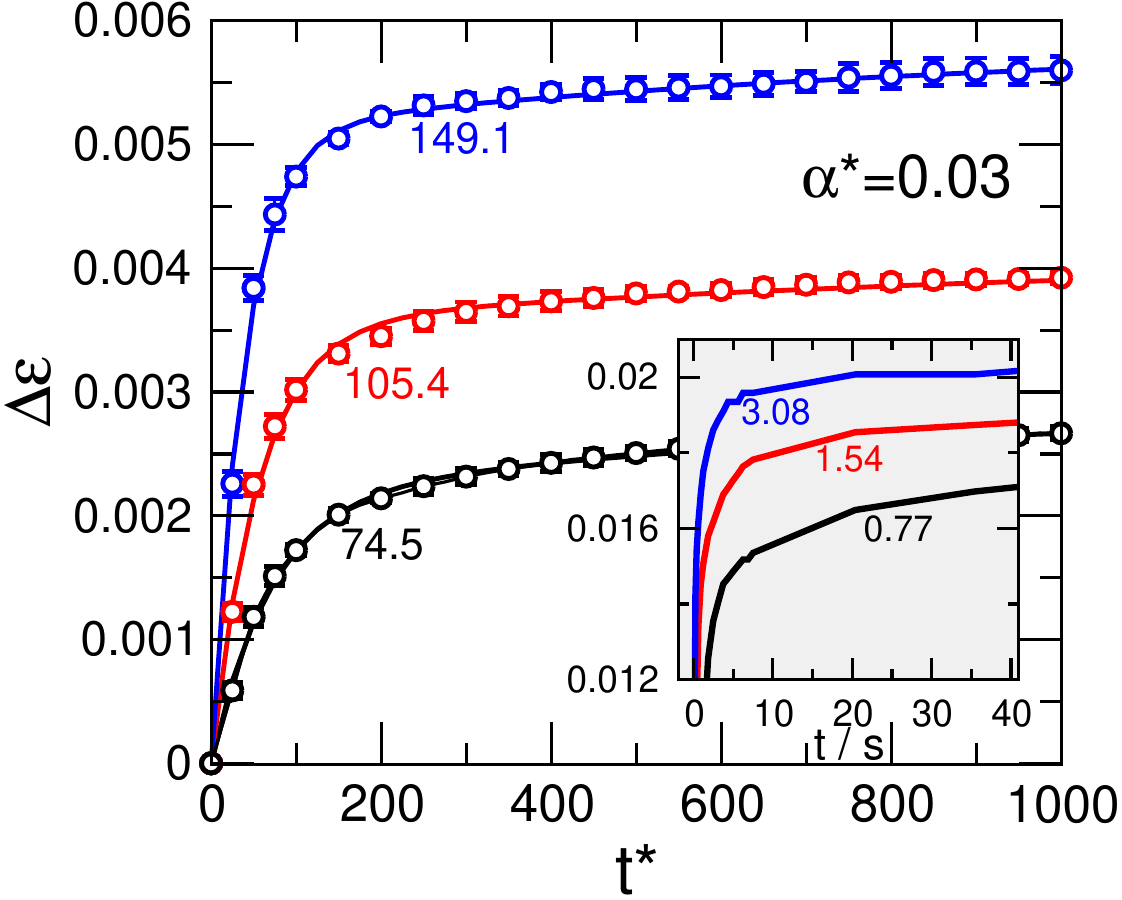}
	\caption{Time dependence of $ \Delta \epsilon$ for $\alpha^{*}=0.03$.
	{The symbols with error bars are simulated data. The error bars have been computed from the variance of the data in the consecutive and independent periods.
	The lines are bi-exponential fits (Eq.~\ref{eq:biexp}).}
	The numbers near the curves indicate the values of the reduced electric field, $E^{*}$. The inset shows the experimental data of Horv\'ath and Szalai~\cite{horvath_pre_2012} for comparison. The numbers in the inset indicate electric field strengths in MV/m unit.  }
	\label{fig:deps_vs_t}
\end{figure}

Fig.~\ref{fig:deps_vs_t} shows the time dependence of $\Delta \epsilon$ for $\alpha^{*}=0.03$ (the curves for other values of $\alpha^{*}$ are similar). 
The numbers near the curves indicate the values of the reduced electric field, $E^{*}$.
This figure corresponds to Fig.~3 of Horváth and Szalai~\cite{horvath_pre_2012} whose data are reproduced in the inset of Fig.~\ref{fig:deps_vs_t} for qualitative comparison.

\begin{figure}[t]
	\centering
	\includegraphics[width=0.4\textwidth]{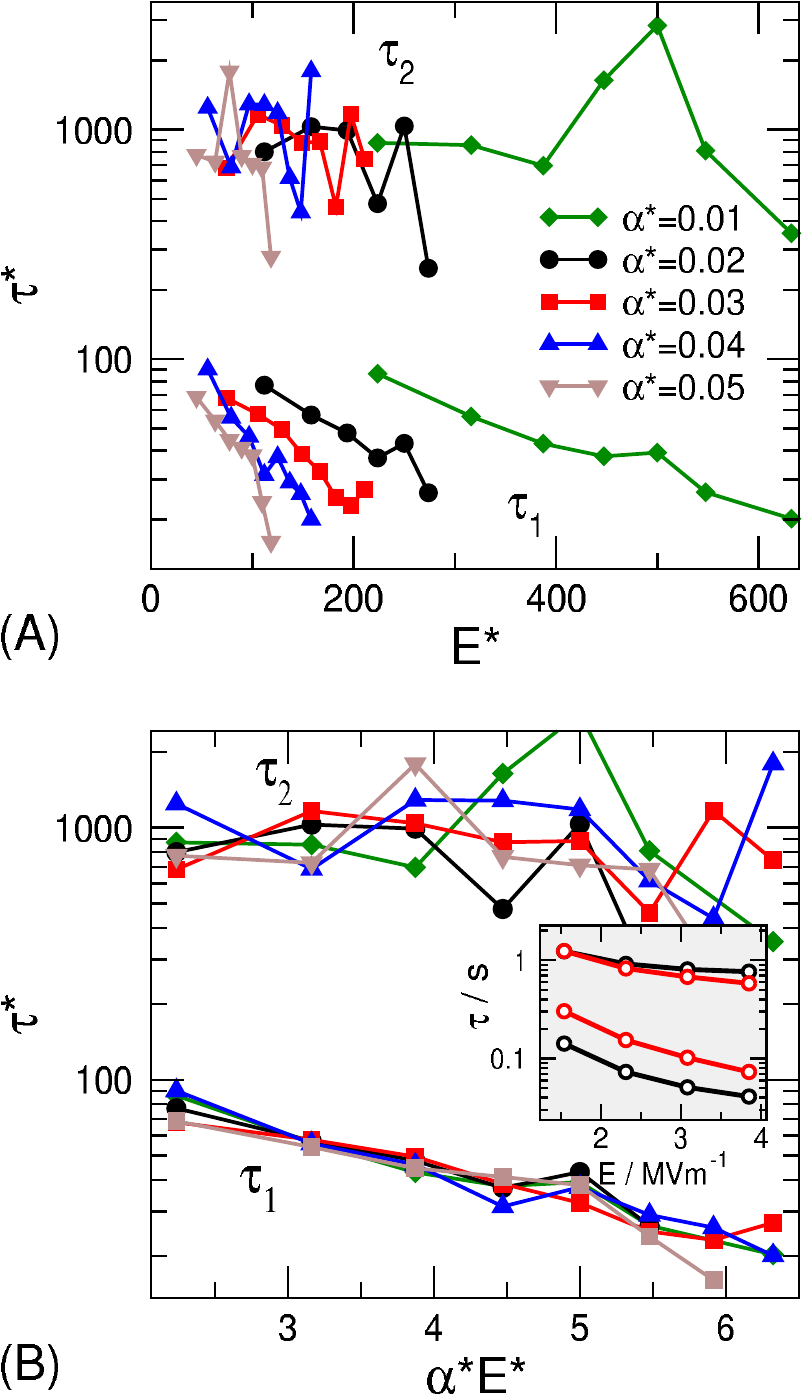}
	\caption{The characteristic times $\tau_{1}$ and $\tau_{2}$ (the latter is larger with an order of magnitude) as functions of (A) $E^{*}$ and (B) $\alpha^{*}E^{*}$ for various values of $\alpha^{*}$. The inset shows the experimental data of Horv\'ath and Szalai~\cite{horvath_pre_2012} (from their Table I.) for the two different values of the viscosity they considered {(black and red colors refer to $\eta=0.34$ and $0.97$ Pa$\,$s, respectively).
	The error bars of $\tau_{1}$ and $\tau_{2}$ estimated from the Levenberg-Marquardt algorithm~\cite{marquardt_1963} are within the size of the sysmbols.}
	}
	\label{fig:tau_vs_E}
\end{figure}

Direct quantitative comparison with the experimental data is problematic (see the end of this section) due mainly to the fact that the ER fluid studied by Horváth and Szalai is polydisperse.
They considered nanosized ($10-20$ nm) silica (SiO$_{2}$) particles dispersed in silicone oils (polydimethylsiloxane) with different dynamic viscosities ($0.34$ and $0.97$ Pa$\,$s).
Qualitative comparison, however, is possible.
By fitting a bi-exponential to our simulated data (Eq.~\ref{eq:biexp}), we obtain the characteristic times (in reduced units) shown in Fig.~\ref{fig:tau_vs_E}.
Panel A shows the data as functions of $E^{*}$ for different values of $\alpha^{*}$.

This figure implies that a larger electric field is needed to achieve smaller $\tau^{*}$ values (faster processes) in the case of smaller $\alpha^{*}$ values.
This is obvious, because the important parameter from the point of view of the dipolar interactions is the induced dipole that is $\alpha^{*}E^{*}$.
Therefore, if we plot $\tau_{1}^{*}$ and $\tau_{2}^{*}$ as functions of $\alpha^{*}E^{*}$, we obtain a scaling behavior: the curves for different $\alpha^{*}$ values collapse onto a single curve (Fig.~\ref{fig:tau_vs_E}B).
{This scaling behavior also applies for the time dependence of the normalized $\Delta \epsilon$; if we plot $\Delta \epsilon(t^{*})/\Delta \epsilon(t^{*}\rightarrow\infty)$ as a function of $t^{*}$ for a fixed value of $\alpha^{*}E^{*}$ but for different combinations of $\alpha^{*}$ and $E^{*}$, the curves collapse onto a single one.}

This figure also shows that $\tau_{2}^{*}$ is an order of magnitude larger than $\tau_{1}^{*}$ in agreement with the predictions of Hass et al.~\cite{hass_pre_1993} as well as with the experiments of Ly et al.~\cite{ly_ijmpb_2001} and Horváth and Szalai.~\cite{horvath_pre_2012}

The qualitative behavior of the $\tau_{1}$ vs.\ $E$ function is also similar to the experimental behavior as seen from comparison to the data in Table I of Horváth and Szalai~\cite{horvath_pre_2012} that are reproduced in the inset of Fig.~\ref{fig:tau_vs_E}B.

\begin{figure*}[t]
	\centering
	\includegraphics[width=0.7\textwidth]{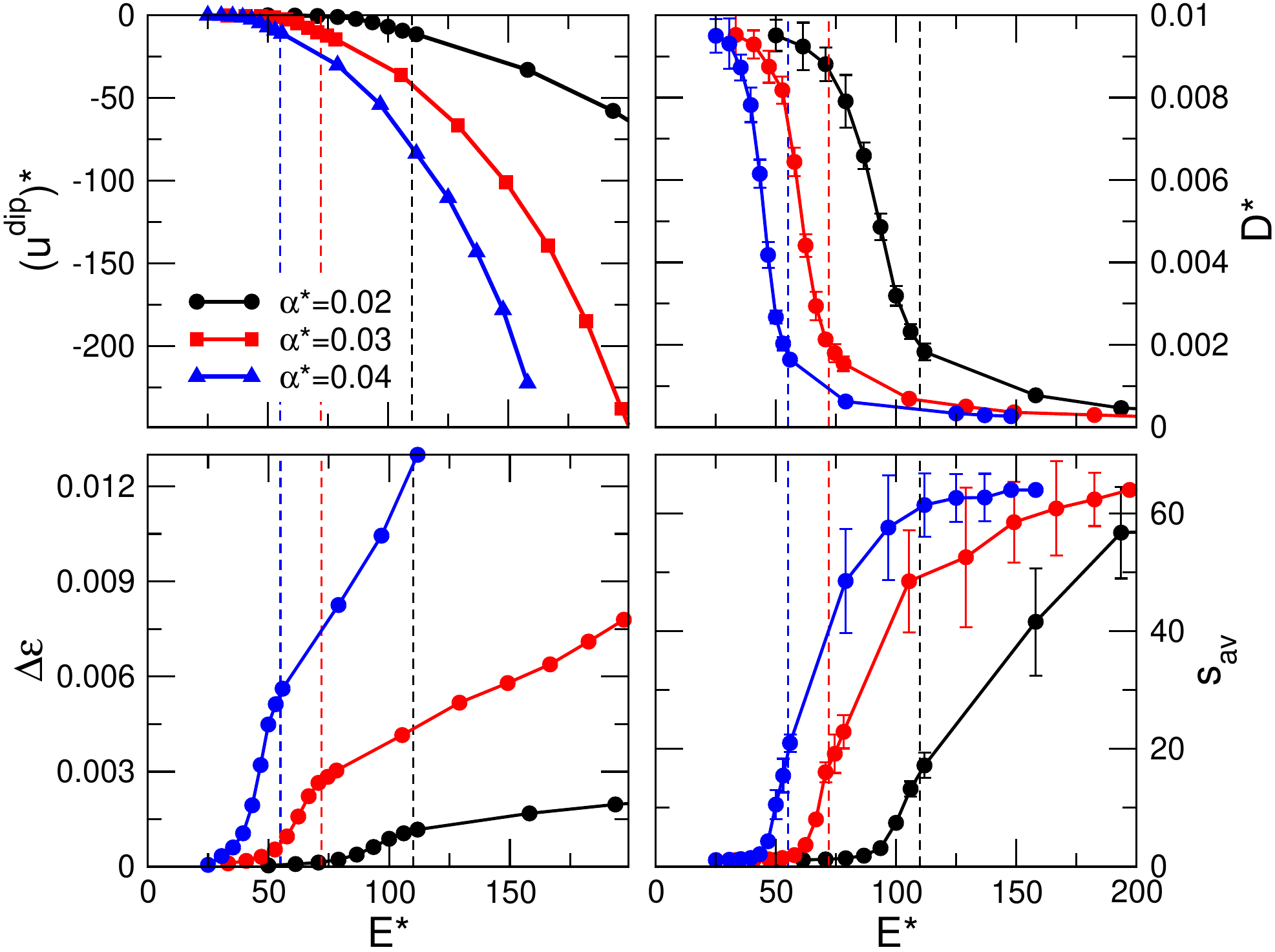}
	\caption{The equilibrium ($t\rightarrow\infty$) values of the one-particle dipolar energy, $(u^{\mathrm{dip}})^{*}$ (top-left panel), the diffusion constant, $D^{*}$ (top-right panel), the induced permittivity increment, $\Delta \epsilon$ (bottom-left panel), and the average chain length, $s_{\mathrm{av}}$ (bottom-right panel), for various values of $\alpha^{*}$. The vertical dashed lines indicate the $E_{\mathrm{h}}^{*}$ field strengths at which the regions with different slopes meet (see Fig.~\ref{fig:deps-E-dep} for more explanation). {The error bars are those computed at $t^{*}=4500$ from the variance of the values in the periods.  The error bars for $(u^{\mathrm{dip}})^{*}$ and $\Delta \epsilon$ are within the size of the symbols.}} 
	\label{fig:E-dep}
\end{figure*}

The purpose of this study is to look into the black box and to see how the dielectric behavior observed in experiment and simulation is related to the structural changes occurring in the ER fluid as the electric field is increased.
These structural changes can be monitored via various physical quantities such as the diffusion constant, the dipolar energy, and the average chain length.
Therefore, we plot the equilibrium values of these quantities (along with $\Delta \epsilon$) as functions of the field strength, $E^{*}$, in Fig.~\ref{fig:E-dep}.
The equilibrium value of a quantity can be obtained either by running a long simulation and throwing the equilibration period away or from substituting $t \rightarrow  \infty$ into the bi-exponential. 
We have chosen the second option.
{Fig.~\ref{fig:deps_vs_t} shows the fitted functions.
The $R^{2}$ coefficient of the fitting is generally above $0.99$ and the residuals (data not shown) decrease to zero as $t\rightarrow\infty$.
These imply that the equilibrium values obtained this way are trustable.}

\begin{figure}[b!]
	\centering
	\includegraphics[width=0.45\textwidth]{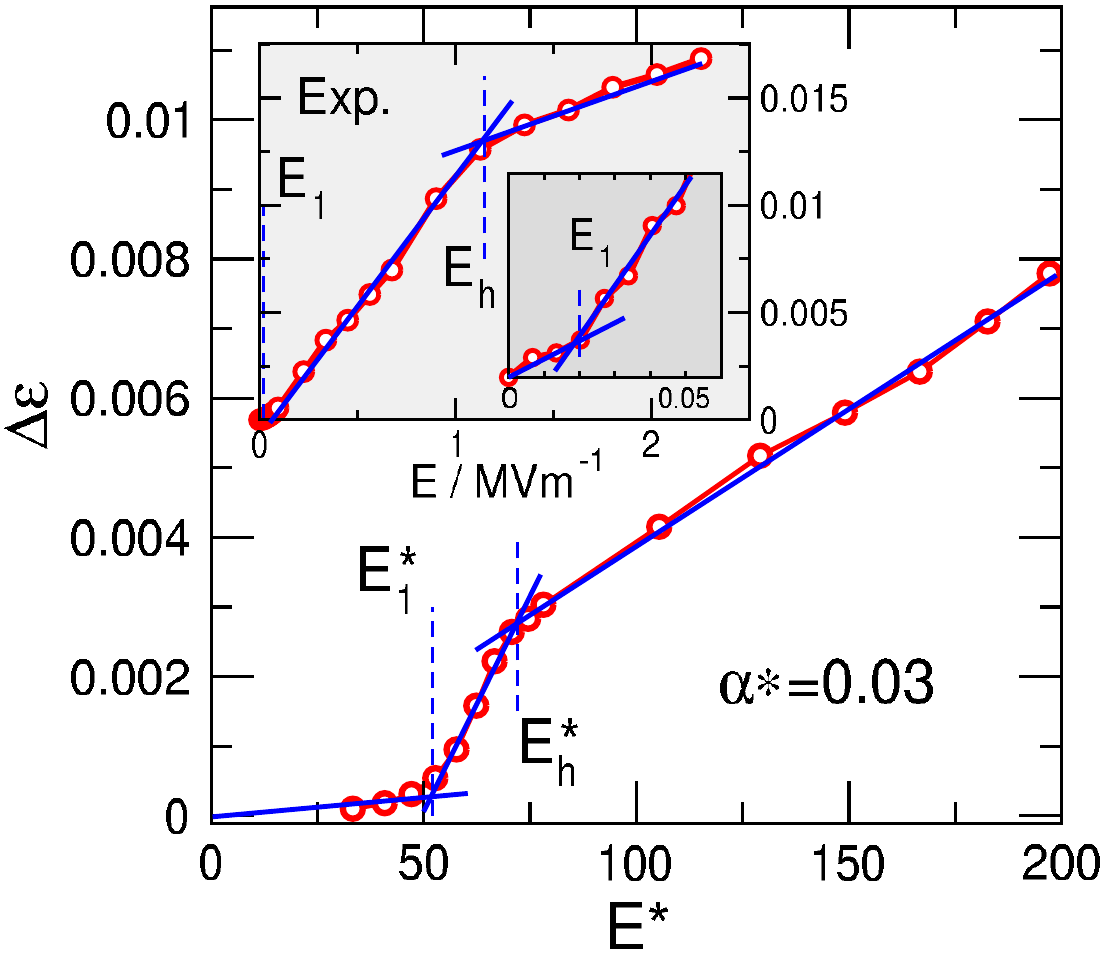}
	\caption{The equilibrium ($t\rightarrow\infty$) value of $\Delta \epsilon$ as a function of $E^{*}$ for $\alpha^{*}=0.03$. The inset shows the experimental data from Fig.~4 of Horv\'ath and Szalai~\cite{horvath_pre_2012}. The inset of the inset shows the results for small $E^{*}$. The solid blue lines indicate the slopes, while dashed blue lines indicate the crosses of the solid lines defining the threshold field strengths $E_{1}$ and $E_{\mathrm{h}}$.}
	\label{fig:deps-E-dep}
\end{figure}

The equilibrium values of $\Delta \epsilon$ as functions of $E^{*}$ are shown in the bottom-left panel of Fig.~\ref{fig:E-dep}.
Also, the curve for $\alpha^{*}=0.03$ is reproduced in Fig.~\ref{fig:deps-E-dep} along with the experimental data in the insets.
The qualitative agreement of the simulations and experimental data is apparent.
The behavior of our curve follows the behavior of the experimental curve in the respect that there is a initial stage with a small slope.
When the electric field is larger than a threshold value ($E_{1}$), pairs and short chains start to form and the curve in the interval between $E_{1}$ and the next threshold value $E_{\mathrm{h}}$ has a larger slope.
Above $E_{\mathrm{h}}$, the chains {spanning} the simulation cell start to aggregate with a smaller slope of the $\Delta \epsilon$ vs.\ $E^{*}$ function.

One notable difference between the simulation and experimental results is that $E_{1}$ is much smaller relative to $E_{\mathrm{h}}$ in the experiment than in the model.
This is valid for all $\alpha^{*}$ values studied.
The reason of this is not clear, but the results may be system-size and density dependent.
{It is also possible that the large-particle fraction of the experimental polydisperse system can form clusters at lower fields than the average-sized particles, an effect that is absent in our monodisperse model.}
The $E_{\mathrm{h}}$ value, however, appears to be a relatively well defined point separating two characteristic regions with different slopes, so we indicate the $E^{*}_{\mathrm{h}}$ vales with vertical dashed lines in all the panels of Fig.~\ref{fig:E-dep}.

{Note that $\Delta \epsilon (E^{*})$ function exhibits a steep nonlinear increase at large $E^{*}$ values.
This is the result of the strong dipolar attraction overriding the repulsion of the WCA potential.
The overlapping spheres lead to stronger particle-particle polarization.
We consider this behavior an artifact of the model.}

Next, let us see how the behavior of the other physical quantities correlates with the behavior of $\Delta\epsilon$.
The one-particle dipolar energy, $(u^{\mathrm{dip}})^{*}$, does not exhibit the behavior of the three (small-large-small) slopes (top-left panel of Fig.~\ref{fig:E-dep}).
It decreases with increasing $E^{*}$ at a continuously increasing rate.
The explanation is that the dipolar energy chiefly depends on the interactions inside the chains.
If we increase $E^{*}$, the dipoles become larger, and also their interactions.
Clustering of chains does not seriously influence this dependence.

The diffusion constant is calculated as the slope of the mean square displacement (MSD) as a function of time: 
\begin{equation}
D(t_{\mathrm{b}})= \frac{\langle \mathbf{r}^{2}(t)\rangle_{\mathrm{b}}}{6 \Delta t_{\mathrm{b}}},
\end{equation} 
where $t_{\mathrm{b}}$ is the time at the beginning of a block, and $\Delta t_{\mathrm{b}}{=}M_{\mathrm{b}}\Delta t$ is the length of the block.
This way, $D$ is characteristic of a block and time dependence can be studied.
$D^{*}$ decreases as $E^{*}$ increases as shown in Fig.~\ref{fig:E-dep} (top-right panel).
The ER particles lose their mobilities as they are organized into chain-like and columnar structures.
The behavior of $D^{*}$ follows the behavior of $\Delta \epsilon$.
Its value starts with the $E^{*}\rightarrow 0$ limit ($0.01$), breaks down around $E_{1}$, and decreases steeply as longer chains are formed at higher $E^{*}$ values.
Around $E_{\mathrm{h}}$, the chains aggregate, $D^{*}$ decreases at a lower rate, and saturates into a very small, but non-zero value.
At large $E^{*}$, the spheres move together with their chains that have much smaller mobility than the single spheres.

{These results are closely related to the anomalous diffusion behavior of dipolar chains that has been studied theoretically and experimentally.~\cite{furst_pre_2000,toussaint_prl_2004,toussaint_pre_2006}
Those studies imply that the diffusion of chains is reduced compared to the case in the absence of chains.
The degree of reduction is related to the average length of the chains.
}

\begin{figure*}[t!]
	\centering
	\includegraphics[width=0.9\textwidth]{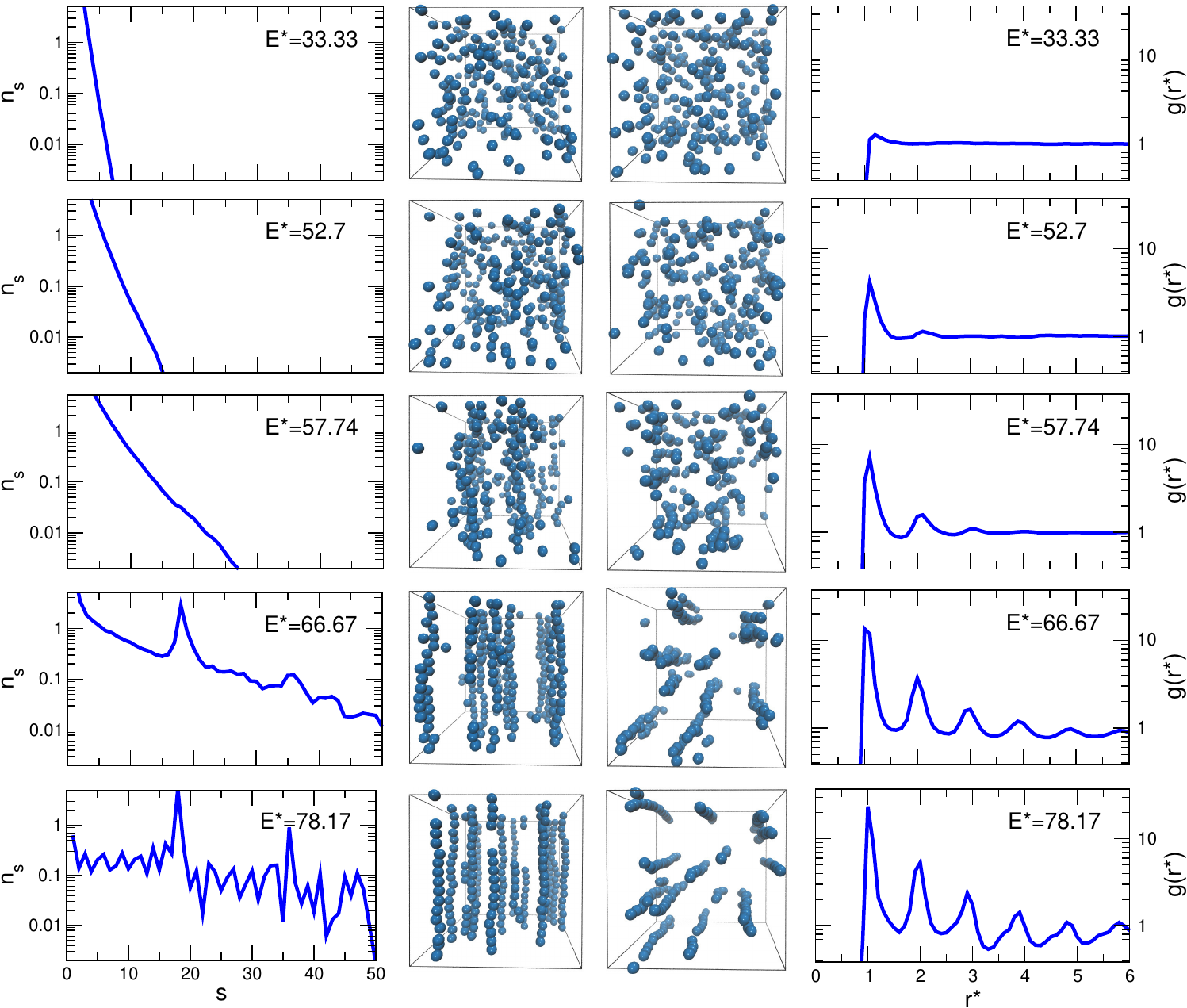}
	\caption{Equilibrium limits of the chain length distribution, $n_{s}$ (left), and radial distribution functions, $g(r^{*})$ (right) for $\alpha^{*}=0.03$. The curves are obtained by averaging over $10$ blocks at the end of $M_{E}$ simulation periods and averaging over periods. Snapshots are shown in the middle in front and top view. The electric field strength increases from top to bottom. }
	\label{fig:composite}
\end{figure*}

The average chain length, $s_{\mathrm{av}}$, is computed by identifying the number of chains of length $s$, $n_{s}$, for every configuration, taking the average 
\begin{equation}
  s_{\mathrm{av}} =\frac{\sum_{s}sn_{s}}{\sum_{s}n_{s}},
  \label{eq:l}
 \end{equation} 
and then averaging over configurations. 
{Chain length, $s$, is measured in the number of particles in the chain.
Two particles are defined to be in the same chain if their distance is smaller than $1.2d$.
The choice of $1.2$ does not influence our qualitative conclusions for the dynamics of chain formation.
Other definitions of chains were analyzed in our previous study~\cite{fertig_hjic_2020}}.

The average chain length starts to increase only above the first threshold value, $E_{1}^{*}$.
Above the second threshold value, $E^{*}_{\mathrm{h}}$, the average length of chains reaches the value $s\approx 18$ that corresponds to a chain {spanning} the simulation cell whose {length} is $L\approx 17.23d$.

As the electric field is increased further well above $E^{*}_{\mathrm{h}}$, the average number of chains also increases and eventually reaches a limiting value of 64.
At large electric fields characteristically $4$ clusters of chains are formed each containing on average 64 spheres, but this is just an average.
There is thermal motion, so values different from 64 may occur, but the average seems solid.
Thermal motion at strong field strengths mainly means the translational motion of chains in the lateral ($x,y$) plane and rotation of chains about the axes of the chains. 
Also, the $4$ clusters of chains do not aggregate further.
At high electric fields, the repulsion between these clusters seems to prevent further aggregation.
This finding, however, is density dependent too.

It is of interest of looking into the black box even deeper to see how these average chain length values come about.
It is done in Fig.~\ref{fig:composite} in which we show chain length distributions, $n_{s}$, radial distribution functions, $g(r)$, and snapshots.
The electric field strength increases from top to bottom.
The values are chosen to show the phases in Figs.~\ref{fig:E-dep} and \ref{fig:deps-E-dep}.
We show results for a value below $E^{*}_{1}$, around $E^{*}_{1}$, between $E^{*}_{1}$ and $E^{*}_{\mathrm{h}}$, around $E^{*}_{\mathrm{h}}$, and above $E^{*}_{\mathrm{h}}$.

As $E^{*}$ increases, the chain length distributions show the increased probability of longer chains.
The snapshots clearly show these chains that are also indicated by peaks in the $g(r)$ functions.

For $E^{*}=33.33$ (below $E^{*}_{1}$), the system is practically a homogeneous isotropic gas-like fluid regarding the ER particles.
For $E^{*}=52.7$ (around $E^{*}_{1}$), pair formation  and, to some degree, formation of short chains are present.
For $E^{*}=57.74$ (between $E^{*}_{1}$ and $E^{*}_{\mathrm{h}}$), even longer chains and, accordingly, more peaks in $g(r)$ appear.

When we reach $E^{*}_{\mathrm{h}}$ ($E^{*}=66.67$), chains {spanning} the simulation cell are clearly present indicated by the peak at $s=18$ in the $n_{s}$ function. 
{This chain is more stabilized by the periodic boundary conditions in our cubic simulation cell of length $L\approx 17.23d$ compared to other chains.}
In our previous study,~\cite{fertig_aipadv_2021} we analyzed this behavior in detail.

For even larger electric field strength ($E^{*}=78.17$), we find another peak at $s=36$ that corresponds to two chains stuck together. 
Chains are straighter, and the peaks in $g(r)$ are more pronounced.
Also, the $n_{s}$ function is more noisy than at lower electric field strengths that indicates that the system is ``more frozen'' or ``less fluid''.
The evolution of the system is determined by the movements of the much less mobile chains instead of the movements of individual particles and short chains.

\begin{figure}[t!]
	\centering
	\includegraphics[width=0.5\textwidth]{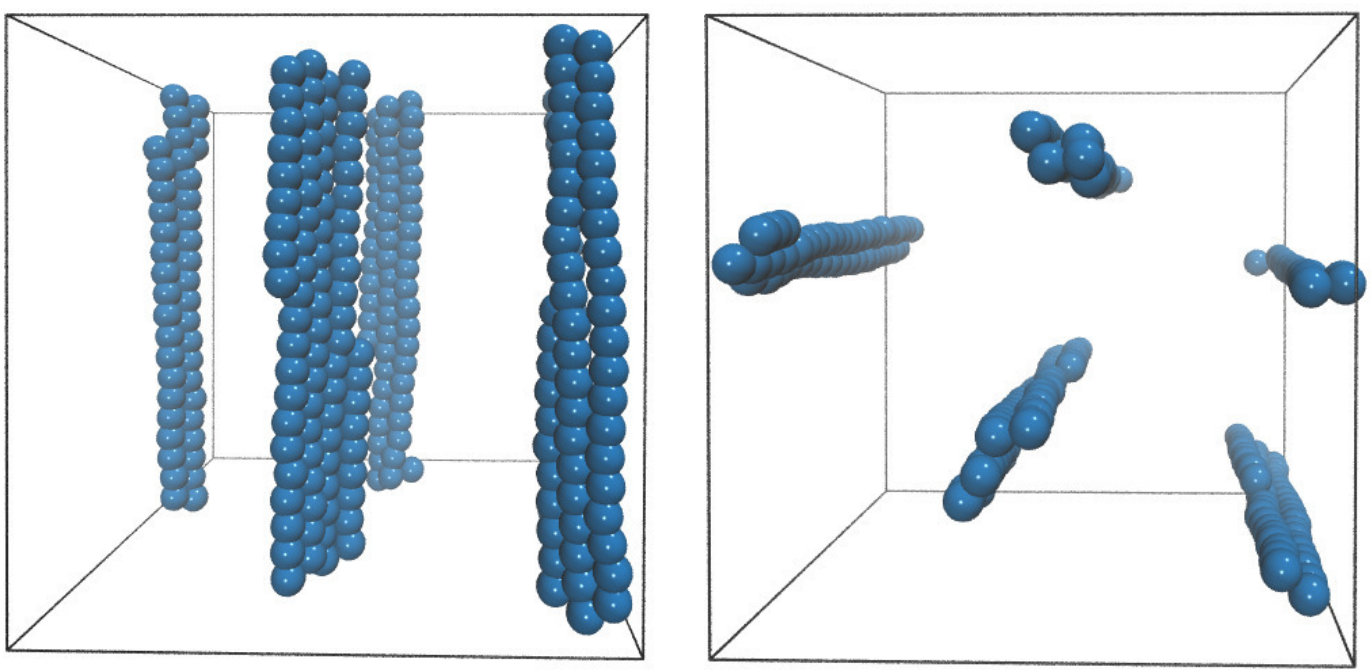}
 	\caption{Snapshots for a very large electric field, $E^{*}=149.07$, for $\alpha^{*}=0.03$.}
    \label{fig:snapshot_columns}
\end{figure}

If we increase the electric field even further ($E^{*}=149.07$), the chains aggregate into column-like structures.
The resulting $n_{s}$ and $g(r)$ profiles are even more noisy and not very meaningful, so we show only snapshots in Fig.~\ref{fig:snapshot_columns}.

Finally, we briefly consider the possibility of a quantitative comparison with the experimental results.
If we accept the hypothesis that the meaning of $E_{\mathrm{h}}$ is the same in the model and in the experiment (it is a cornerstone), we can estimate the particle diameter, $d$, that brings the simulation and the experimental data into correspondence.
If we take the value of $E_{\mathrm{h}}=1.18$ MV/m from Fig.~4 of Ref.~\onlinecite{horvath_pre_2012} and relate it to the $E_{\mathrm{h}}^{*}$ values depicted from Fig.~\ref{fig:E-dep} ($E_{\mathrm{h}}^{*}=52.5$, $72$, $107$ for $\alpha^{*}=0.02$, $0.03$, and $0.04$, respectively), we obtain the values $d=673$, $517$, and $418$ nm for $\alpha^{*}=0.02$, $0.03$, and $0.04$, respectively.
These values are much larger than the $10-20$ nm values specified in the paper of Horv\'ath and Szalai.~\cite{horvath_pre_2012}

We can explain this in different ways.
First, the ER particles provided by the manufacturer are polydisperse as opposed to our model that includes particles of the same size.
Also, it was observed in the experiments that the particles tend to stick together forming larger particles resulting in a larger effective diameter.
Water content that may increase polarizability cannot be excluded.
In any case, the values $10-20$ nm are so small that using them in a simulation (by transforming them to the corresponding reduced units) does not result in any kind of chain formation.
For these reasons, we regard the diameters calculated and reported here more realistic than the $10-20$ nm values.

It is also possible to correspond the $\tau_{1}^{*}$ value depicted from Fig.~\ref{fig:tau_vs_E} ($\tau_{1}^{*}\approx 68.5$ for $\alpha^{*}=0.03$) to the experimental $\tau_{1}$ value depicted from the inset of Fig.~\ref{fig:tau_vs_E}B ($\tau_{1}\approx 0.38$ s for viscosity $0.97$ Pa$\,$s).
The resulting diameter is $d\approx2470$ nm.  
The reason of these large values is that our reduced friction coefficient is small ($\gamma^{*}=100$).
Because we can extrapolate to larger values of $\gamma^{*}$ (see Fig.~6 of our previous work~\cite{fertig_aipadv_2021}), we can provide the estimation that by increasing $\gamma^{*}$ with two orders of magnitude, the resulting diameter decreases with about one order of magnitude.
Changing $\gamma^{*}$ does not change the equilibrium value of $s_{\mathrm{av}}$ and $\Delta \epsilon$, it only influences how fast the system converges to these values.

{To relate $t^{*}$ to $t$ and $E^{*}$ to $E$ in Figs.~\ref{fig:deps_vs_t}, \ref{fig:tau_vs_E} and \ref{fig:deps-E-dep}, we collect the unit values $E_{0}=E/E^{*}$ and $t_{0}=t/t^{*}$ in Table  \ref{tab:Et} for three representative values of the particle diameter: $d=517$ and $2467$ nm obtained from the procedures of relating $E_{\mathrm{h}}$ to $E_{\mathrm{h}}^{*}$ and $\tau_{1}$ to $\tau_{1}^{*}$, respectively, and a value in between ($1000$ nm).}

\begin{table}[t]
\caption{Unit values of the electric field strength and particle diameter, $E_{0}=E/E^{*}$ and $t_{0}=t/t^{*}$, respectively, for different values of $d$ on the basis of Table \ref{tab:reduced} ($T=298.15$ K and $\rho_{\mathrm{in}}=2650$ kg/m$^{3}$).}
\label{tab:Et}
 	\def\arraystretch{1.3}
 	\centering
\begin{tabular}{lll} \hline 
$d$ / nm \hspace{0.3cm} & $E_{0}$ / MVm$^{-1}$ \hspace{0.3cm}& $t_{0}$ / s \\ \hline
$516$ & $0.016$ & $0.00011$ \\
$1000$ & $0.0061$ & $0.00058$ \\
$2467$ & $0.0016$ & $0.0055$ \\ \hline
\end{tabular}
\end{table}

\section{Conclusions}

We performed Brownian Dynamics simulations for ER fluids by taking the cross-polarization among particles into account in a self consistent way and computed the induced dielectric increment, $\Delta  \epsilon$, as a function of time after the applied field is switched on and as a function of field strength at the equilibrium limit.
{Particle-particle polarization is essential for computing $\Delta \epsilon$ that is a very useful quantity because it is measurable well and also can be obtained from simulations with a small statistical error.}

Our results are in qualitative agreement with the experimental results of Horv\'ath and Szalai~\cite{horvath_pre_2012} and relate the computed data to structural features in terms of energy, diffusion constant, average chain length, chain length distribution, radial distribution functions, and snapshots.
The hypotheses about the correlations between dielectric properties and structural features put forward by Horv\'ath and Szalai have been confirmed by our calculations.

\section*{Author's contribution}

All authors contributed equally to this work.

\section*{Acknowledgments}

This research was supported by the European Union, co-financed by the European Social Fund, via the project ``Research of autonomous vehicle systems related to the autonomous test track in Zalaegerszeg'' (EFOP-3.6.2-16-2017-00002).
We also acknowledge the support of the National Research, Development and Innovation Office (NKFIH), project No.~K124353. 
We acknowledge KIFÜ for awarding us access to resource based in Hungary at Szeged.

\section*{Data avalability}

The data that supports the findings of this study are available within the article.

%


\end{document}